\documentstyle[mncite,draft,psfig]{mn}

\title[IR spectroscopy of CVs -- III. DNe below the period gap and novalikes]
{Infrared spectroscopy of cataclysmic variables: III. \newline
Dwarf novae below the period gap and novalike variables}

\author[V.\,S.\ Dhillon et al.]{V.\,S.\ Dhillon$^{1}$, 
S.\,P.\ Littlefair$^{1}$, S.\,B.\ Howell$^{2}$, D.\,R.\ Ciardi$^{3}$,
M.\,K.\,Harrop-Allin$^{4}$ \\
{\normalsize \,\,\ \ \,}\\
{\LARGE and T.\,R.\,Marsh$^{5}$}\\
{\normalsize \,\,\ \ \,}\\
$^1$Department of Physics and Astronomy, University of Sheffield, Sheffield
S3 7RH, UK \\
$^2$Astrophysics Group, Planetary Science Institute, 620 North 6$^{th}$ Ave.,
Tuscon, AZ 85705, USA \\
$^3$Department of Astronomy, 211 Bryant Space Sciences Building, 
University of Florida, Gainesville, FL 32611, USA \\
$^4$Mullard Space Science Laboratory, University College London, 
Holmbury St. Mary, Dorking, Surrey RH5 6NT, UK \\
$^5$Department of Physics and Astronomy, University of Southampton, 
Highfield, Southampton SO17 1BJ, UK
}
\date{\center{\Large Accepted for publication in the Monthly Notices of the
Royal Astronomical Society \\ 
\vspace{.5cm} \today}} 

\begin{document}
\maketitle

\begin{abstract} 
We present K-band spectra of the short-period dwarf novae YZ~Cnc, LY~Hya, 
BK~Lyn, T~Leo, SW~UMa and WZ~Sge, the novalike variables 
DW~UMa, V1315~Aql, RW~Tri, VY~Scl, UU~Aqr and GP~Com, and a 
series of field dwarf stars with spectral types ranging from K2--M6. 

The spectra of the dwarf novae
are dominated by emission lines of H\,{\small I} and 
He\,{\small I}. The large velocity and equivalent widths of
these lines, in conjunction with the fact that the lines are
double-peaked in the highest inclination systems, indicate an
accretion disc origin. 
In the case of YZ Cnc and T Leo, for which we obtained time-resolved
data covering a complete orbital cycle, the emission lines show
modulations in their equivalent widths which are most probably
associated with the bright spot (the region where the gas stream 
collides with the accretion disc).  
There are no clear detections of the secondary star in any of the 
dwarf novae below the period gap, yielding upper limits of
10--30\% for the contribution of the secondary star to the observed
K-band flux. In conjunction with the K-band magnitudes of the 
dwarf novae, we use the derived secondary star contributions to
calculate lower limits to the distances to these systems. 

The spectra of the novalike variables are
dominated by broad, single-peaked emission lines of H\,{\small I} and 
He\,{\small I} -- even the eclipsing systems we observed do not 
show the double-peaked profiles predicted by standard accretion disc theory. 
With the exception of RW~Tri, which exhibits Na\,{\small I}, 
Ca\,{\small I} and $^{12}$CO absorption features consistent with a 
M0V secondary contributing 65\% of the observed K-band flux, we 
find no evidence for the secondary star in any of the novalike variables. 
The implications of this result are discussed. 

\end{abstract} 


\section{Introduction}
\label{sec:introduction}
\newcommand{\gta}{{\small\raisebox{-0.6ex}
{$\,\stackrel{\raisebox{-.2ex}{$\textstyle >$}}{\sim}\,$}}}
\newcommand{\lta}{{\small\raisebox{-0.6ex}
{$\,\stackrel{\raisebox{-.2ex}{$\textstyle <$}}{\sim}\,$}}}

Cataclysmic variables (CVs) are interacting binary systems in which a
white dwarf primary accretes material from a red dwarf secondary (see
\pcite{warner95a} for a review). The infrared (IR) is a relatively 
unexplored part of the spectrum as far as CVs are concerned (see 
\pcite{dhillon97a} for a review). This is suprising given the fact that the 
1--2.5 $\mu$m wavelength range happens to be where spectrum of the 
G--M-dwarf secondary star in CVs is expected to 
peak\footnote{The spectral type of the secondary star in a CV can be
determined from its  orbital period, $P$, using the relationships
$26.5-0.7 P$ (for  $P<4$~hr) and $33.2-2.5 P$ (for $P>4$~hr), where
G0=0, K0=10 and M0=20 \cite{smith98}. The wavelength,
$\lambda_{max}$, of the peak flux, $f_{\nu}$, for  a star of effective
temperature $T_{e\!f\!f}$ can be approximated by:
$\lambda_{max}=5100/T_{e\!f\!f}$~$\mu$m. With $T_{e\!f\!f}$ ranging
from $\sim$6000--2000~K for the G--M-dwarf secondary stars in CVs,
$\lambda_{max}$ ranges from $\sim$1--2.5~$\mu$m.},
and where low-harmonic, cyclotron emission from the weaker-field magnetic CVs
and emission from  the cool, outer  regions of  the  accretion disc in
non-magnetic CVs would be expected to  fall. For the above reasons, 
we embarked on a spectral survey of CVs in the IR, the results of which
have been published in Dhillon \&\ Marsh (1995: 
Paper I -- dwarf novae above the
period gap) and Dhillon et al. (1997: Paper II -- intermediate polars). 

Two of the types of CV which were not surveyed in Papers I and II were the
dwarf novae below the period gap and the novalike variables. These represent 
the two classes of CV in which the secondary star has proved to be hardest to
detect in optical spectra (e.g. \pcite{friend88}, \pcite{smith97}). 
In fact, table 1 of \scite{smith98} shows that the secondary star
has been directly observed in only 4 dwarf novae below the period gap and
4 novalike variables, compared to a total of 55 detections in all
classes of CV. In the case of dwarf novae below the period gap, 
the paucity of secondary star detections is probably due to the 
fact that the secondary star is of a very late-type and hence faint. In 
the case of the novalike variables, the lack of secondary star detections
is probably due to the very bright discs drowning out the light from the 
companion. If we are to learn anything about the secondary stars in these
systems, and hence how these types of CV evolve, some means of 
detecting the secondary star must be found. Given that light from the
disc should be relatively weak in the IR, and that the secondary star
should peak in the IR, we decided to perform an IR spectral survey of 
dwarf novae below the period gap and novalike variables in order to 
detect their elusive secondary stars. The results of this survey are
presented in this third and final paper in the series.

\section{Observations}
\label{sec:observations} 

The data presented in this paper were obtained between 
1993 February 8 and 1997 May 28 with the 
1--5~$\mu$m Cooled Grating Spectrometer 4
(CGS4) on the United Kingdom 3.8~m Infrared Telescope (UKIRT) on Mauna Kea, 
Hawaii. With the exception of GP~Com, which was observed with the
instrumental configuration described by \scite{dhillon97}, all of the 
spectra presented in this paper were obtained with the 
256$\times$256 pixel InSb array, the 75~l/mm grating in first order
and the 150~mm camera, giving a wavelength range of approximately 
0.6~$\mu$m at a resolution of 350~km\,s$^{-1}$. 
Optimum spectral sampling and bad pixel 
removal were obtained by mechanically shifting the array over 2 pixels in 
the dispersion direction in steps of 0.5 pixels.
We employed the non-destructive readout mode of the detector 
in order to reduce
the readout noise. The slit width was 1.2 arcseconds (projecting 
to 1 pixel on the detector) and the slit 
was oriented at the parallactic angle. The
seeing disc was usually equal to or slightly larger in size than the 
slit width, except on the night of 1997 May 28, when we employed the 
new tip-tilt secondary mirror to produce sub-arcsecond images. The
observations were all obtained in photometric conditions, apart from
the night of 1995 October 21 which suffered from a little high cirrus. 
In order to compensate for fluctuating atmospheric 
OH$^-$ emission lines \cite{ramsay92} we took relatively short exposures 
and nodded the telescope primary so that the object spectrum switched between 
two different spatial positions on the detector. A full journal of 
observations is presented in table~\ref{tab:journal}. 

\begin{table*}
\caption[]{Journal of observations. The classifications and orbital 
periods of the CVs have been taken from the catalogue of \scite{ritter98}.
The spectral types of the late-type dwarfs have been taken from the 
catalogue of \scite{kirkpatrick91}, unless otherwise noted.
Note that the classification of BK Lyn as the first novalike variable below
the period gap \cite{dobrzycka92} has been questioned by \scite{ringwald96},
who speculate that the object may instead be a dwarf nova
with rare outbursts akin to those observed in objects like WZ Sge. 
For the purposes of this paper, we have grouped BK Lyn with the dwarf novae
below the period gap.}
\boldmath{
\begin{center}
{\normalsize\bf
\begin{tabular}{llccccc}
& & & & & & \\
\multicolumn{7}{c}{Cataclysmic variables} \\
& & & & & & \\
& & & & & & \\
\multicolumn{1}{l}{Object} &
\multicolumn{1}{l}{Class} &
\multicolumn{1}{c}{Period} &
\multicolumn{1}{c}{Date} &
\multicolumn{1}{c}{UTC} &
\multicolumn{1}{c}{UTC} &
\multicolumn{1}{c}{Exposure time} \\
& & \multicolumn{1}{c}{(hours)} & & \multicolumn{1}{c}{start} &
\multicolumn{1}{c}{end} & \multicolumn{1}{c}{(seconds)} \\
& & & & & & \\
RW Tri        & NL UX       & 5.57 & 21/10/95 & 08:30 & 09:05 & \ 1700 \\
VY Scl        & NL VY       & 3.99 & 21/10/95 & 07:37 & 08:18 & \ 1900 \\ 
UU Aqr        & NL UX       & 3.93 & 21/10/95 & 09:16 & 09:32 & \ \ \,800 \\
V1315 Aql     & NL UX SW    & 3.35 & 08/09/95 & 07:54 & 08:05 & \ \ \,480 \\
DW UMa        & NL UX SW    & 3.28 & 03/02/97 & 11:14 & 11:53 & \ 1920 \\
BK Lyn        & NL SH?      & 1.80 & 06/02/96 & 10:18 & 11:14 & \ 2880 \\
GP Com        & NL AC       & 0.78 & 08/02/93 & 15:39 & 16:10 & \ 1440 \\ 
YZ Cnc        & DN SU       & 2.08 & 05/02/96 & 07:20 & 11:34 & 12000 \\
LY Hya        & DN          & 1.80 & 05/02/96 & 13:14 & 15:32 & \ 6720 \\
T Leo         & DN SU       & 1.41 & 06/02/96 & 13:22 & 15:25 & \ 5760 \\
SW UMa        & DN SU DQ?   & 1.36 & 06/02/96 & 07:17 & 10:11 & \ 8160 \\
WZ Sge        & DN SU WZ CP & 1.36 & 28/05/97 & 11:49 & 14:39 & \ 8640 \\
& & & & & & \\
\end{tabular}
\begin{tabular}{llcccc}
& & & & & \\
\multicolumn{6}{c}{Late-type dwarfs} \\
& & & & & \\
& & & & & \\
\multicolumn{1}{l}{Object} &
\multicolumn{1}{l}{Spectral type} &
\multicolumn{1}{c}{Date} &
\multicolumn{1}{c}{UTC} &
\multicolumn{1}{c}{UTC} &
\multicolumn{1}{c}{Exposure time} \\
& & & \multicolumn{1}{c}{start} &
\multicolumn{1}{c}{end} & \multicolumn{1}{c}{(seconds)} \\
& & & & & \\
GL764.1A     & K2V$^a$   & 20/10/95 & 07:26 & 07:35 & 192 \\
GL775        & K5V       & 20/10/95 & 06:20 & 06:27 & \,96 \\ 
GL764.1B     & K7V       & 20/10/95 & 05:43 & 05:52 & 240 \\ 
GL154        & M0V$^b$   & 21/10/95 & 09:39 & 09:48 & 192 \\
GL763        & M0V       & 20/10/95 & 06:00 & 06:08 & 144 \\
GL229        & M1V       & 06/02/96 & 05:00 & 05:08 & \,72 \\
GL806        & M2V       & 20/10/95 & 07:12 & 07:21 & 192 \\
GL436        & M3V       & 05/02/96 & 11:56 & 12:06 & 192 \\
GL748AB      & M3.5V     & 28/05/97 & 11:17 & 11:26 & 576 \\
GL402        & M4V       & 05/02/96 & 11:41 & 11:46 & 120 \\
GL866AB      & M5V       & 21/10/95 & 05:25 & 05:33 & 192 \\
GL473AB      & M5.5V$^c$ & 06/02/96 & 15:44 & 16:02 & 288 \\
GL65AB       & M6V       & 05/02/96 & 05:14 & 05:22 & 192 \\
& & & & & \\
& & & & & \\
\multicolumn{6}{l}{$^a$\scite{gliese69}} \\
\multicolumn{6}{l}{$^b$\scite{hawley96}} \\
\multicolumn{6}{l}{$^c$\scite{henry94}} \\
& & & & & \\
\end{tabular}
}
\end{center}
}
\label{tab:journal}
\end{table*}

\section{Data Reduction} 
\label{sec:datared} 

The initial steps in the reduction of the 2D frames were performed
automatically by the CGS4 data reduction system \cite{daley94}. 
These were: the application of
the bad pixel mask, bias and dark frame subtraction, flat field division,
interlacing integrations taken at different detector positions, and co-adding
and subtracting nodded frames. Further details of the above procedures may be
found in the review by \scite{joyce92}. In order to obtain
1D data, we subtracted the residual sky and then optimally
extracted the spectra \cite{horne86a}. 

There were three stages to the calibration of the spectra. The first was the
calibration of the wavelength scale using argon arc-lamp exposures. The
second-order polynomial fits to the arc lines always yielded an error of less
than 0.0003~$\mu$m (rms). The next step was the removal of the ripple arising
from variations in the star brightness between integrations (i.e. at different
detector positions). These variations were due to changes in the seeing, sky
transparency and the slight motion of the stellar image relative to the slit.
We discovered that the amplitude of the ripple varied across the spectrum and
we were thus forced to apply a correction for this based on a linear
interpolation of the ripple profiles at either end of the spectrum. The final
step in the spectral calibration was the removal of telluric atmospheric
features and flux calibration. This was performed by dividing the spectra to be
calibrated by the spectrum of an F-type standard, observed at a similar
airmass 
(typically within 0.1), with its prominent stellar features masked out.
We then multiplied the result by the known flux of the standard at each
wavelength, determined using a black body function set to the same temperature
and magnitude as the standard. As well as providing flux
calibrated spectra, this procedure also removed atmospheric absorption features
from the object spectra. 

\section{Results}
\label{sec:results}

\begin{figure*}
\centerline{\psfig{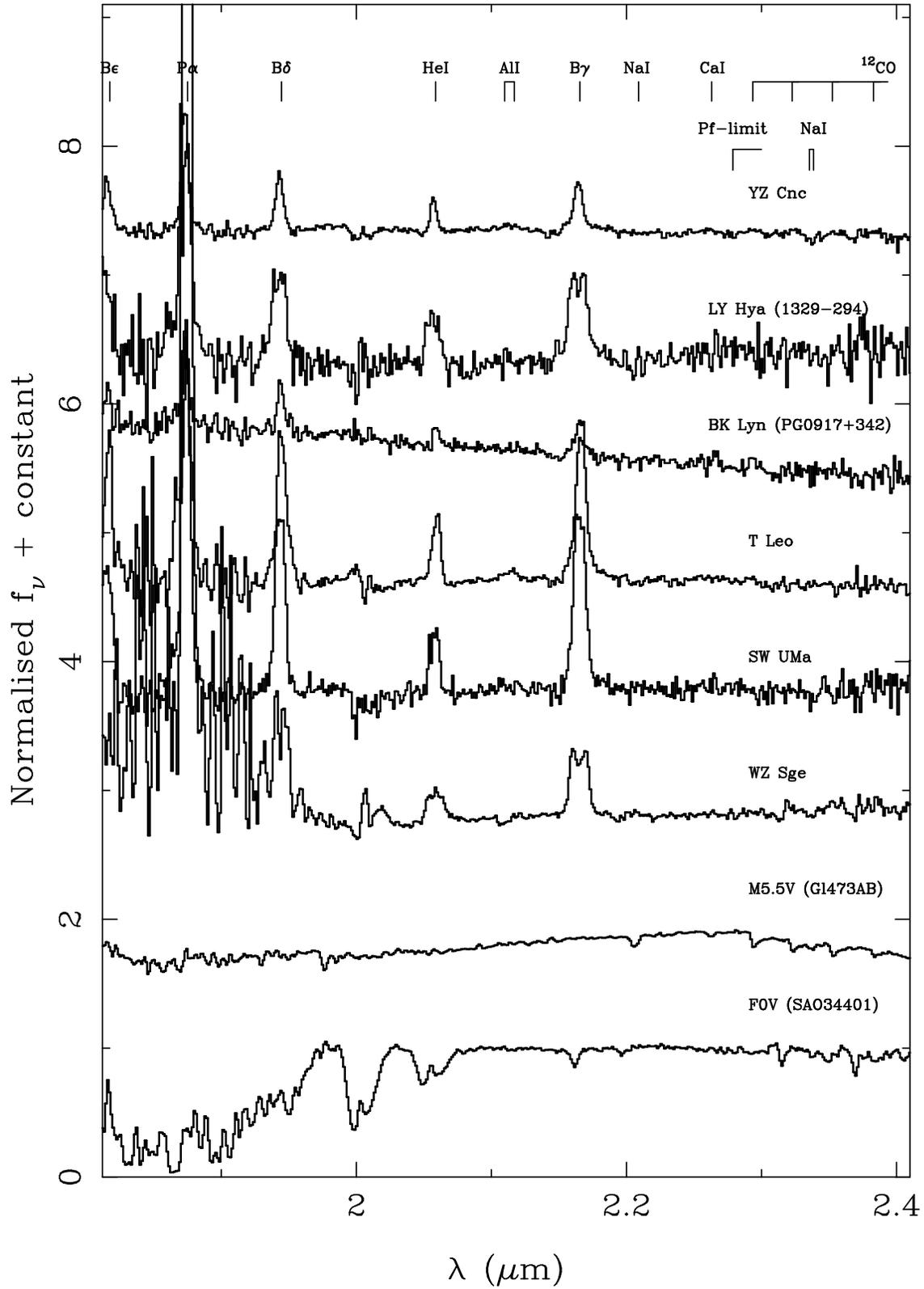}}
\caption{K-band spectra of the dwarf novae
YZ~Cnc, LY~Hya, BK~Lyn, T~Leo, SW~UMa, WZ~Sge and an M5.5V star.
The spectra have been normalised by dividing by the flux 
at 2.24 $\mu$m and then offset by adding a multiple of 0.9
to each spectrum. Also shown is the spectrum of an F0V star,
normalised by dividing by a spline fit to its continuum, which
indicates the location of telluric absorption features; spectral features 
within the strongest absorption bands are highly uncertain.} 
\label{fig:shortdne}
\end{figure*}

\begin{figure*}
\centerline{\psfig{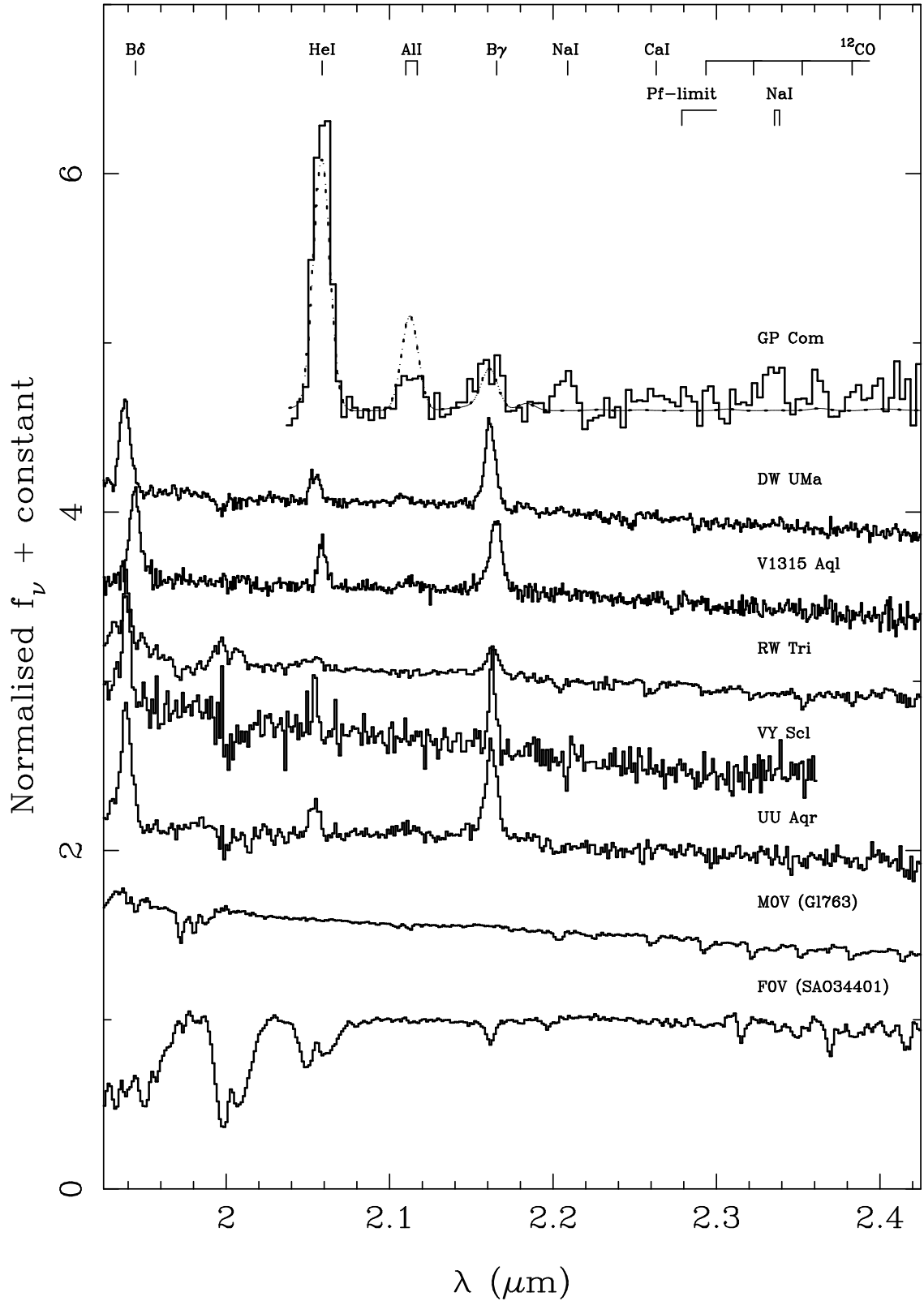}}
\caption{K-band spectra of the novalike variables GP~Com, 
DW~UMa, V1315~Aql, RW~Tri, VY~Scl, UU~Aqr and an M0V star.
The spectra have been normalised by dividing by the flux 
at 2.24 $\mu$m and then offset by adding a multiple of 0.5
to each spectrum. Also shown is the spectrum of an F0V star,
normalised by dividing by a spline fit to its continuum, which
indicates the location of telluric absorption features; spectral features 
within the strongest absorption bands are highly uncertain. The
dashed line under the spectrum of GP~Com is a model spectrum
from gas in LTE (see section~\protect\ref{sec:emlines} for details).}
\label{fig:nmnl}
\end{figure*}

\begin{figure*}
\centerline{\psfig{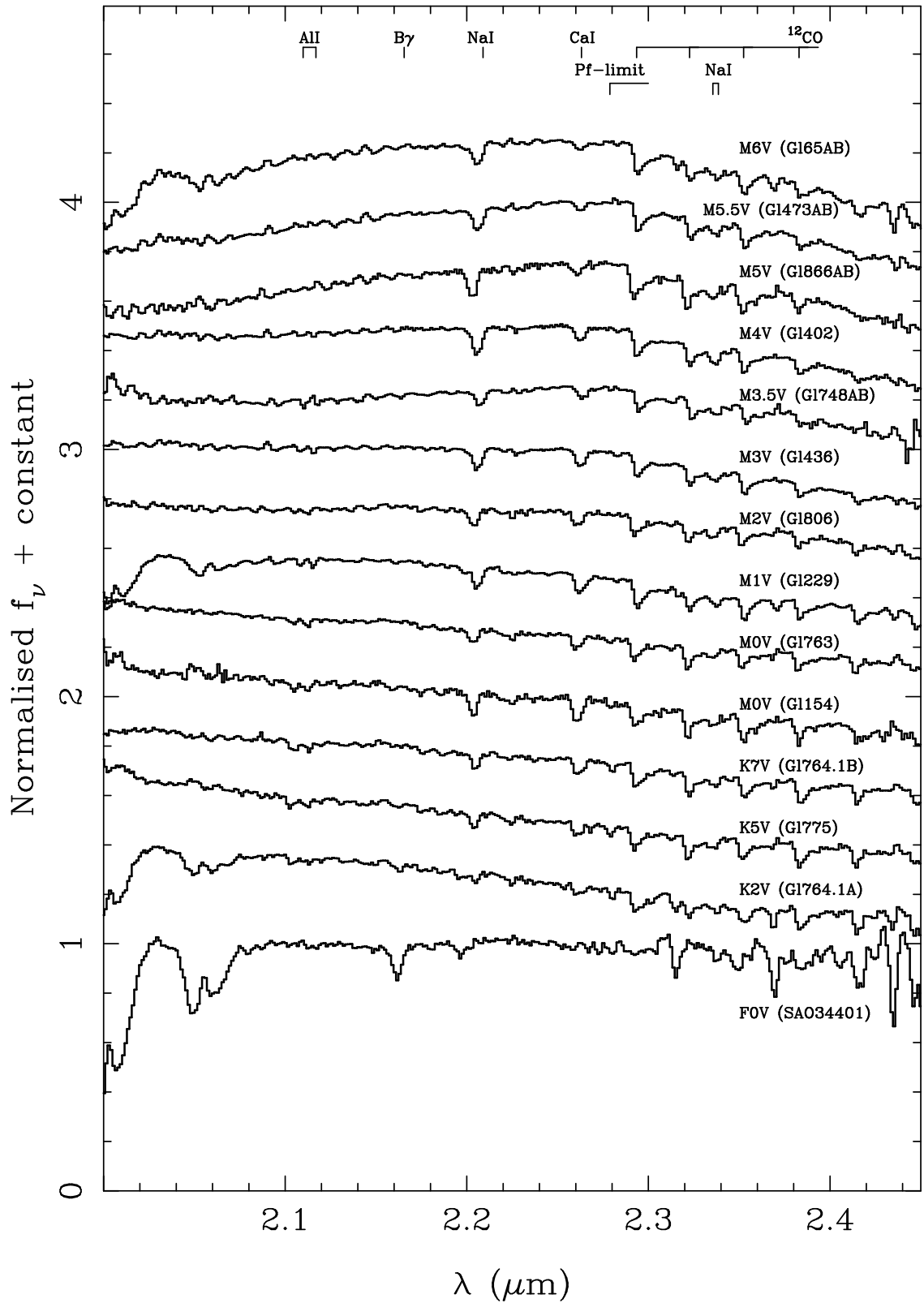}}
\caption{K-band spectra of K2V--M6V spectral-type template stars.  
The spectra have been normalised by dividing by the flux 
at 2.24 $\mu$m and then offset by adding a multiple of 0.25
to each spectrum. Also shown is the spectrum of an F0V star,
normalised by dividing by a spline fit to its continuum, which
indicates the location of telluric absorption features; spectral features 
within the strongest absorption bands are highly uncertain.} 
\label{fig:sptypes}
\end{figure*}

Figure~\ref{fig:shortdne} shows the K-band spectra
of the dwarf novae YZ~Cnc, LY~Hya, BK~Lyn, T~Leo, SW~UMa and WZ~Sge, and
figure~\ref{fig:nmnl} shows the K-band spectra
of the novalike variables GP~Com, DW~UMa, V1315~Aql, RW~Tri, 
VY~Scl and UU~Aqr. Note that the anti-dwarf novae DW UMa and VY Scl were
both in their high state when the spectra in figure~\ref{fig:nmnl} were
obtained. In figure~\ref{fig:sptypes} we show the K-band spectra
of main-sequence field dwarfs ranging in spectral 
type from K2 to M6 and in tables~\ref{tab:lines}~and~\ref{tab:seclines}
we list the wavelengths, equivalent widths 
and velocity widths of the most 
prominent spectral lines identified in 
figures~\ref{fig:shortdne},~\ref{fig:nmnl}~and~\ref{fig:sptypes}. 

\subsection{Emission lines}
\label{sec:emlines}

The spectra of the dwarf novae below the period gap in
figure~\ref{fig:shortdne} are dominated  by strong
emission lines  of He\,{\small I}  and the Paschen and Brackett
series of  H\,{\small I}.  The large velocity and equivalent widths of 
these lines (see table~\ref{tab:lines}) indicate an accretion disc
origin. This conclusion is further supported by the double-peaked
emission-line profiles exhibited by the high-inclination dwarf novae
LY Hya and WZ Sge (see also \pcite{skidmore99} and \pcite{mason99}) in 
figure~\ref{fig:shortdne}. 

With the exception of the double-degenerate system GP Com, which will be
discussed in more detail below, the spectra of the novalike variables
in figure~\ref{fig:nmnl} are dominated
by strong, single-peaked  emission lines of H\,{\small  I} (Brackett-$\gamma$
and  Brackett-$\delta$) 
and  He\,{\small I}  (2.0587~$\mu$m).  This is  in stark
contrast to what one might expect from standard accretion disc theory
(e.g. \pcite{horne86}),
which  predicts that emission  lines from high inclination discs, such
as  those in the  eclipsing systems DW UMa, V1315   Aql, RW Tri and UU
Aqr,    should appear double   peaked.  The  absence of  double-peaked
profiles in high-inclination novalikes is also observed in the optical
(see \pcite{dhillon96}) and is one of  the defining 
characteristics of the so-called SW Sex stars, of which DW UMa, 
V1315 Aql and UU Aqr are members. 

GP Com consists of a CO white dwarf and a helium degenerate star in an
orbit  of 46~min  period. Neither  star   is directly  visible  -- the
optical  light  from the system  is dominated  by the  accretion disc,
which has  a spectrum composed almost entirely  of helium and nitrogen
emission lines  reflecting  the products of  hydrogen burning  and CNO
processing in the helium degenerate donor. A  very simple model, based
upon LTE   emission from an  $\sim11\,000$~K  optically  thin (in the
continuum)  slab,  provides a  surprisingly   good fit to  the optical
spectrum of GP Com \cite{marsh91}. In figure~\ref{fig:nmnl}, the same
model has been applied to the IR spectrum of GP Com, with equally good
results. The  model  predicts the existence  of three  strong emission
lines in the K-band, all  of He\,{\small I},  which are all present in
the  actual spectrum. Note  that  there  is also some  evidence  for a
fourth emission line in the  spectrum at 2.2~$\mu$m, apparently unrelated
to any telluric features, that we have been unable to identify. 

\subsection{Absorption lines/bands} 
\label{sec:absn}

RW Tri is the only CV presented in  
figures~\ref{fig:shortdne}~and~\ref{fig:nmnl} which
shows absorption  features from the  secondary star --  one can
clearly make out the  profiles of Al\,{\small I}, Na\,{\small I},
Ca\,{\small   I} and $^{12}$CO in the     spectrum.     The
distinctive    water   absorption   band at $\sim$2.3~$\mu$m, so
prominent in the spectra of the late M-stars presented in
figure~\ref{fig:sptypes},  is absent in RW Tri, indicating that its
secondary is most likely a late K-dwarf. 
In order to estimate the spectral type
of the secondary star in RW~Tri from our data, and to determine its
contribution to the total K-band flux, we used an optimal subtraction
technique (e.g. \pcite{dhillon93}). First, we normalised the spectra
of RW~Tri and the  template stars by dividing by a third-order
polynomial fit to the continuum.  A constant times the normalised
template spectrum was then  subtracted from the normalised spectrum of
RW~Tri and the constant adjusted so as to minimise the residual
scatter in regions containing secondary star features.  The residual
scatter is measured by carrying out the subtraction and then computing
the $\chi^2$-value  between the residual  spectrum and a smoothed
version of itself. Prior to the subtraction,  the template spectra
should be broadened to account for the rotational  velocity of the
secondary star; the  low resolution of the our data, however, made
this step unnecessary. The value obtained for the percentage contribution
naturally depends on spectral type -- the correct spectral type being the
one which minimises the value of $\chi^2$. Using this method we find that
the secondary star contributes $65\pm5\%$ (1-$\sigma$ error) of 
the K-band light in RW~Tri
and that the spectral type of the secondary star is M0V. 
This spectral-type determination is in good agreement with the
results of the skew mapping experiments of \scite{smith93b},
who derive a spectral type of K7V or cooler, and with 
the predictions of the main-sequence spectral type-period
relationship of \scite{smith98}, which suggests a spectral type of
K9V. 

None of the other CVs show any signs of absorption features 
from the secondary star. In these cases it
is possible to determine an upper limit to the contribution of the secondary
star. This is done by subtracting a constant times the normalised template
spectrum from the normalised CV spectrum, until spectral absorption features
from the template star appear in emission in the CV spectrum. The value of
the constant at this point represents an upper limit to the fractional
contribution of the secondary star. The contribution found depends on the
spectral type of the template used. As the spectral type of the secondary star
is unkown for these systems, we used the period-spectral type relationship
of \scite{smith98} to select an appropriate spectral type for the template 
star. The spectral type used for each CV, along with the corresponding 
upper limits to the secondary star contribution, 
are listed in table~\ref{tab:contrib}.

It is seen that the secondary star contributes less than 30-55\% for 
the novalikes and 10-30\% for the dwarf novae below the period gap.
Exceptions are VY~Scl and LY~Hya, where poor signal-to-noise has resulted
in very high values -- the significance of these results is that we would have
to obtain better signal-to-noise if we are to have any hope of observing
the secondary in these systems, regardless of its contribution to the
K-band light. A further exception is WZ~Sge, in which the secondary must
contribute less than 10\% of the K-band light. This is in agreement with
\scite{littlefair99} who find that the secondary star in WZ~Sge contributes
less than 20\% to the J-band light, and \scite{ciardi98} who model the 
near-IR flux in WZ~Sge and find that the secondary is a cool 
($\sim$1700 K) star which contributes approximately 10\% of the overal IR
flux. 

\subsection{Distances}

The distances to CVs can be measured from K-band spectra using a modification
of a method first proposed by \scite{bailey81}. The distance modulus can
be rewritten in terms of the K-band surface brightness as
\begin{equation}
S_k = m_{K}+5-5 \log d + 5\log(R/R_{\odot}),
\label{eq:dist}
\end{equation}
where $m_K$ is the apparent K-band 
magnitude, $d$ is the distance in parsecs and $R$
is the radius of the star. For a star of one solar radius, $S_k$ is equal to
the absolute K-band magnitude. Since the radius of the secondary star is equal
to the radius of the
Roche lobe, the orbital period and mass of the secondary are sufficient
to determine its radius (there is also a weak dependance on mass ratio).
$m_K$ is derived from the K-band magnitude of the CV
and the percentage contribution of the secondary star. 
Given $m_K$ and the value of
$S_k$, which can be obtained from the $V-K$ colour (or spectral type) of
the secondary using the empirical calibrations derived from field dwarfs by
\scite{ramseyer94}, it is possible to estimate the distance using the above 
equation. Note that the distances to only 5 of the surveyed CVs are
estimated here, as the remaining 7 CVs do not have published K-band
magnitudes.

\subsubsection{RW~Tri}
The K-band magnitude of RW~Tri is 11.59 \cite{longmore81}. In conjunction
with the percentage contribution estimated in section~\ref{sec:absn}, 
this gives a K-band magnitude for the secondary star of $m_K=12.06\pm0.08$. 
Assuming the error in $S_k$ is dominated by the error in spectral type 
(estimated to be one spectral type sub-classification), \scite{bessel91} 
gives a $V-K$ colour for the secondary of $V-K =3.65\pm0.1$. Using the 
calibrations in \scite{ramseyer94} this gives $S_k = 4.43\pm0.1$. 
We have estimated the radius of the secondary to be $R/R_{\odot} = 
(0.61 \pm 0.2)$ from the orbital period-radius relation (equation 11) given by
\scite{smith98}. These values, in conjuction with equation~\ref{eq:dist} give
a distance to RW~Tri of $d=205\pm90$ parsecs. \scite{mcarthur99} used the 
Hubble Space Telescope to obtain a parallax for RW~Tri, establishing the
distance to RW~Tri at $341^{-31}_{+38}$ parsecs. The discrepancy in the two 
results may be due to the fact that the actual K-band 
magnitude of RW~Tri at the time of our observation is unknown and we have 
had to adopt the value obtained by \scite{longmore81}. We know, 
however, that there is a long-term variation in the $V$-band magnitude of 
RW~Tri at primary minimum, ranging from 13.7 to 15.4, suggesting there
is an uneclipsed, variable component \cite{longmore81}. One might expect to
see variation on a similar scale in the K-band (the actual long-term K-band 
variation has never been recorded), in which case our results are 
consistent with distances ranging from 90--400 parsecs.

\subsubsection{WZ Sge}
The K magnitude of WZ~Sge is 13.3 \cite{ciardi98}, which in conjunction with
our upper limit to the secondary star contribution implies that
$m_K \geq 15.8$. \scite{smak93} 
estimates the radius of the secondary to be $R_2/R_{\odot} = 0.11$.
Determining $S_k$ is more problematic. \scite{ciardi98} find evidence that
the secondary in WZ~Sge is very cool (less than 1700 K) and \scite{littlefair99}
use distance estimates by \scite{spruit98} and \scite{smak93} to constrain
the spectral type of the secondary to be later than M7.5. $S_k$ for
M7.5 (from \pcite{bessel91}) is 6.8 and we shall adopt this value. 
Using these values we obtain a distance to WZ~Sge of $d \geq 69$ parsecs. 
\scite{spruit98} and \scite{smak93} both found a 
distance to WZ~Sge of 48 parsecs by fitting white dwarf models to UV 
spectra of WZ Sge obtained using the HST and the IUE, respectively.

\subsubsection{T Leo}
The K magnitude of T~Leo is 14.01 \cite{sproats96}, which implies $m_K \geq
15.3$. We have estimated the radius of the secondary from the orbital 
period-radius relation given by \scite{smith98}. 
Likewise, $S_k$ follows
from the data in \scite{bessel91} and a spectral type estimated from the
orbital period-spectral
type relationship in \scite{smith98}. Using these values
we find $d \geq 120$ parsecs, close to the value of $d \sim 100$ parsecs
found by \scite{howell99}.

\subsubsection{YZ Cnc}
The K magnitude of YZ~Cnc is 14.37 \cite{sherrington83}, which implies
$m_K \geq 16$. Again, we estimate $S_k$ and the radius of the secondary star
from the orbital period and the relationships given by \scite{smith98}.
We find $d \geq 290$ parsecs. \scite{patterson84} finds a distance to YZ~Cnc
of $\sim 130$ parsecs, using a method based on the equivalent width of
H$\beta$ and the shape of the continuum. The large discrepancy between
these numbers is almost certainly due to the relatively unreliable 
techniques which \scite{patterson84} was forced to use.

\subsubsection{BK Lyn}
The K magnitude of BK~Lyn is 14.63 \cite{sproats96}. Following the same 
method as for YZ~Cnc and T~Leo, we obtain $d \geq 185$ parsecs. Note that
\scite{dobrzycka92} derived a value of $d \sim 1000$ parsecs, but this was
based upon the assumption that the absolute magnitude of BK~Lyn is
the same as that typically found for novalikes (i.e. $M_V \sim +4$), and
is hence highly uncertain. 

\subsection{Time resolved spectra}
Time resolved spectra were obtained for T~Leo, SW~UMa, YZ~Cnc
and WZ Sge. 
Unfortunately, a combination of poor signal-to-noise and velocity resolution
means that radial velocity studies and Doppler tomography were fruitless
(with the exception of WZ Sge -- see \pcite{skidmore99} and
\pcite{mason99}).
Skew mapping (see \pcite{smith93b}) was performed, but no systems showed
signs of the secondary star. Equivalent width light curves were obtained
for all stars, but only those of T~Leo and YZ~Cnc show significant features,
and are described here.

\subsubsection{YZ Cnc}
\begin{figure*}
\centerline{\psfig{figure=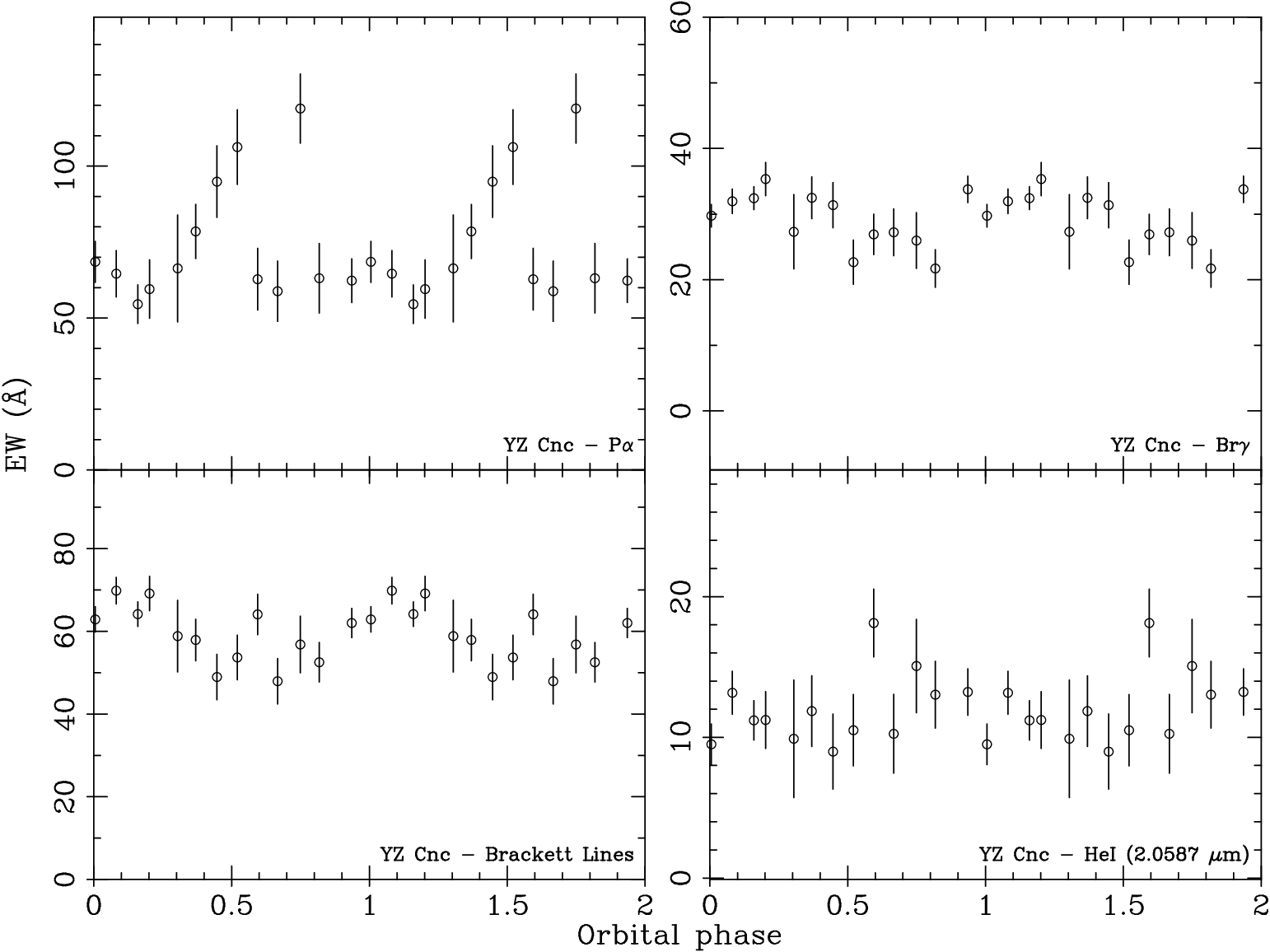,width=15cm,angle=-90}}
\caption{Equivalent width (EW) 
light curves of YZ Cnc. The light curves have been folded over 
two orbital cycles for clarity, 
according to the ephemeris of \protect\scite{vanparadijs94}.
Paschen-$\alpha$ was determined in the range
1.854--1.890 $\mu$m, Brackett-$\gamma$ in the range 2.153--2.173 $\mu$m, 
He\,{\small I} in the range 2.053--2.062 $\mu$m and the
sum of the Brackett-$\gamma$ and Brackett-$\delta$ lines in the range 
1.933--1.952 and 2.153--2.173 $\mu$m.}
\label{fig:yzcnclight}
\end{figure*}

Figure~\ref{fig:yzcnclight} shows the equivalent width
light curves for YZ~Cnc. Both the
Brackett lines and He\,{\small I} show a modulation on the orbital period, with
maximum light at $\phi \sim 0.0$.  
\scite{vanparadijs94} find a similar modulation in the optical data, but with
different phasing, with maximum light at $\phi \sim 0.8$. They 
attribute this to bright spot modulations. It is likely that the modulations
we observe in the K-band
share the same source, given that the discrepancy in phase
is well within the error quoted for the ephemeris given by
 \scite{vanparadijs94}.
The variation is not as marked in the K-band as in the optical, perhaps
sugesting that the bright spot contributes relatively less light to the 
system in the IR. 

The Paschen-$\alpha$ line in YZ~Cnc shows unusual behaviour, but this
line is strongly affected by telluric features and hence the data are
uncertain.

\subsubsection{T Leo}
\begin{figure*}
\centerline{\psfig{figure=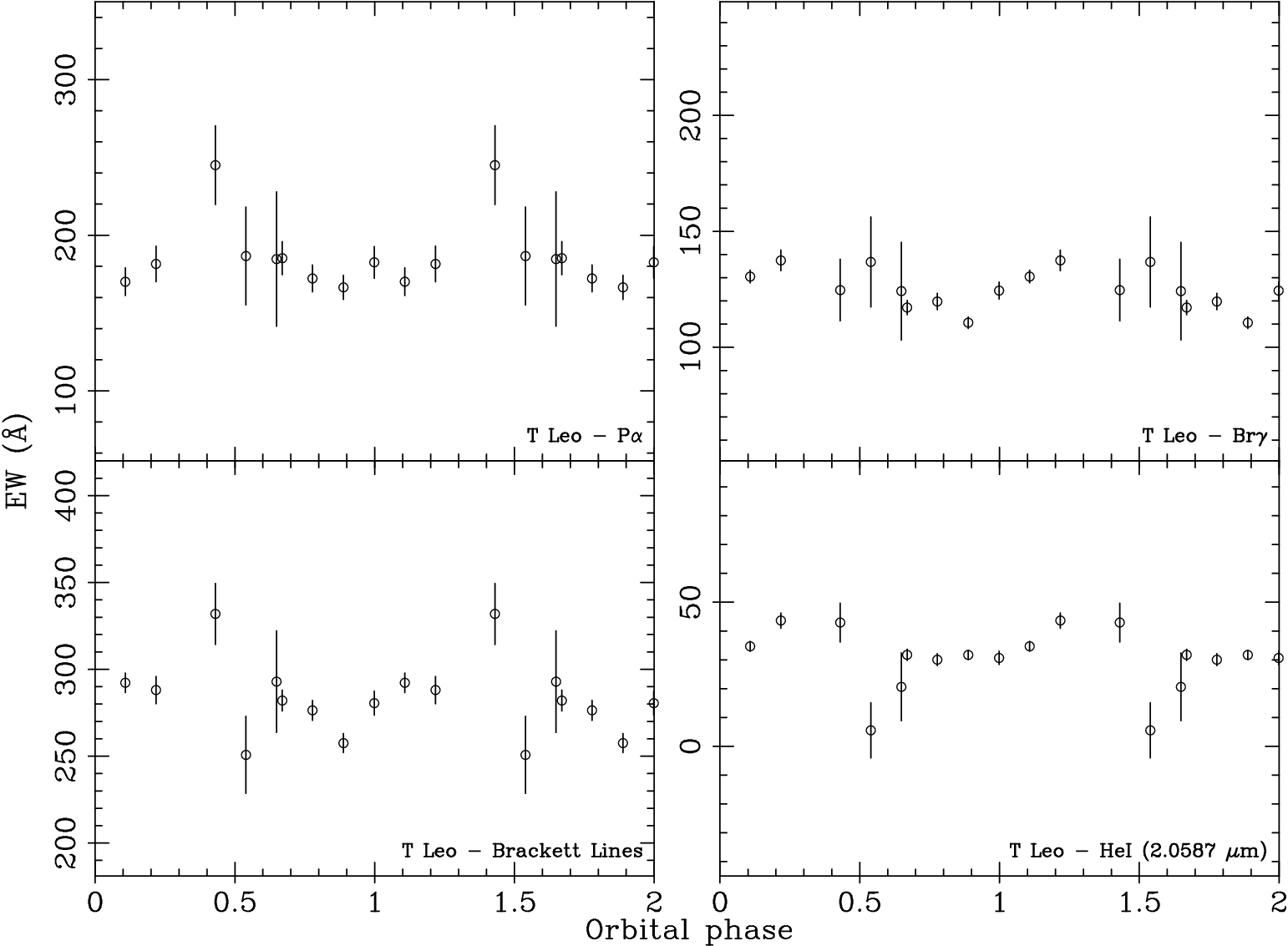,width=15cm,angle=-90}}
\caption{Equivalent width (EW) 
light curves of T Leo. The light curves have been folded over 
two orbital cycles for clarity, 
according to the ephemeris of \protect\scite{shafter84b}.
Paschen-$\alpha$ was determined in the range
1.854--1.890 $\mu$m, Brackett-$\gamma$ in the range 2.153--2.173 $\mu$m, 
He\,{\small I} in the range 2.053--2.062 $\mu$m and the
sum of the Brackett-$\gamma$ and Brackett-$\delta$ lines in the range 
1.933--1.952 and 2.153--2.173 $\mu$m.}
\label{fig:tleolight}
\end{figure*}

Figure~\ref{fig:tleolight} shows the equivalent width 
light curves for T~Leo. The data is of
poorer time resolution and signal-to-noise than the YZ~Cnc data of
figure~\ref{fig:yzcnclight}. Paschen-$\alpha$ and
the Brackett lines appear to show a similar variation to that observed
 in YZ~Cnc, with the exception that 
maximum light is now located at $\phi \sim 0.4$. Phase should not be taken as
reliable in this case as a considerable time has elapsed since the ephemeris
of \protect\scite{shafter84b} was determined. 
We suggest that this variation may possibly be due to bright
spot modulations, as in YZ~Cnc, but poor signal-to-noise and errors
in phasing make this conclusion uncertain.
The He\,{\small I} line does not show clear modulations with orbital phase.

\section{Discussion} 
\label{sec:discussion} 

The prospects of studying the secondary stars in novalike
variables (particularly the SW Sex-type systems just above the
period gap) and dwarf novae below the period gap appears bleak. 
With a few rare exceptions, the 
secondary star cannot even be detected, let alone studied. 
This, in turn, means that an essential tool in the study of the 
evolution of these systems -- a knowledge of the mass and spectral 
type of the secondary star -- is currently unavailable.

Why are the secondary stars so difficult to detect in these systems?
In the case of the dwarf novae below the period gap, this is almost
certainly due to the low surface temperature and small size of the 
secondary star, with the secondary contributing no more
than 10--30\% of the total K-band light. This compares with the
easily detectable secondary stars in dwarf novae above the period
gap, which typically contribute at least 75\% of the K-band light
\cite{dhillon95a}.

Novalike variables also lie above the period gap, however, so why
are their secondary stars not visible? One explanation is that the
steady-state accretion discs in novalike variables are 
much brighter than the quiescent discs in dwarf novae of a similar 
orbital period; the resulting increase in shot noise from the disc 
spectrum overwhelms the weak signal from the secondary star. This
does not explain, however, why the secondary star in DW UMa was not
even visible during a 3--4 magnitude low-state (when accretion had 
virtually ceased) in the I-band spectra of \scite{marsh97}. This 
observation implies 
that the secondary star in DW UMa has an apparent magnitude of
I$>$19.5 and hence a distance of at least $\sim$850 pc if the 
secondary star has spectral type M4.$^*$  If this lower-limit to the 
distance is typical of most novalikes (or specifically, SW Sex stars), 
then it means that the mass transfer rates derived from techniques 
such as eclipse mapping (e.g. \pcite{rutten92b}) are underestimating the true
values. 

\section*{\sc Acknowledgements}

We would like to thank Simon Duck and Lee Sproats for their help with
putting together the various observing proposals which led to the award
of telescope time for this survey.
We would also like to thank Tariq Shahbaz for his comments on this paper.
UKIRT is operated by the Joint Astronomy 
Centre on behalf of the Particle Physics and Astronomy Research Council.
The spectra of V1315~Aql and DW~UMa were obtained through the UKIRT service 
programme. The data reduction and analysis were performed at the Sheffield
node of the UK STARLINK computer network. SBH would like to acknowledge
partial support of this work by NASA grants NAG5-4233 and GFSC-070.

\bibliographystyle{mnras}
\bibliography{abbrev,refs}

\begin{table*}
\caption[]{Wavelengths, equivalent widths and velocity widths
of the most
prominent lines visible in the IR spectra of the surveyed CVs. 
The line identifications have been based upon the list presented
by \protect\scite{dhillon95a} and references therein. 
The wavelengths given for the $^{12}$CO lines refer to the band-heads. 
The vertical bars following lines of similar wavelength indicate that the 
equivalent width measurements apply to the entire blend.
The two-letter codes indicate 
that a line was either not present (np) or present but not measurable (nm).}

\newcommand{\vd}{$\pm$}
\renewcommand{\arraystretch}{1.0}
\begin{center}
\boldmath{
\hspace*{-7.0cm}
{\scriptsize
\begin{tabular*}{11.0cm}{@{\extracolsep{-3.18mm}}llccccccccccccccc}
 & & & & & & & & & & & & & & & & \\
 &
&\multicolumn{3}{c}{\bf VY Scl}&\multicolumn{3}{c}{\bf RW Tri}
&\multicolumn{3}{c}{\bf UU Aqr}&\multicolumn{3}{c}{\bf LY Hya}
&\multicolumn{3}{c}{\bf YZ Cnc}\\
{\small\bf Line}&$\lambda$&\bf{EW}&\bf{FWHM}&\bf{FWZI}&\bf{EW}&\bf{FWHM}
&\bf{FWZI}&\bf{EW}&\bf{FWHM}&\bf{FWZI}&\bf{EW}&\bf{FWHM}&\bf{FWZI}&\bf{EW}
&\bf{FWHM}&\bf{FWZI}\\
&{$\mu$\bf{m}}&\bf{\AA}&{\small\bf kms$^{-1}$}&{\small\bf kms$^{-1}$}&\bf{\AA}
&{\small\bf kms$^{-1}$}&{\small\bf kms$^{-1}$}
&\bf{\AA}&{\small\bf kms$^{-1}$}&{\small\bf kms$^{-1}$}&\bf{\AA}
&{\small\bf kms$^{-1}$}&{\small\bf kms$^{-1}$}&\bf{\AA}&{\small\bf kms$^{-1}$}
&{\small\bf kms$^{-1}$}\\
 & & & & & & & & & & & & & & & & \\ 
{\small\bf B-}$\epsilon$  & {\small\bf 1.8174} & nm & nm & nm & nm & nm & nm &nm & nm & nm & nm &3000\vd1000& nm & nm & 1050\vd100 & nm \\
{\small\bf P-}$\alpha$     & {\small\bf 1.8751} &nm &nm &nm &nm &nm &nm &nm &nm&nm &170\vd20 &2200\vd150 &5200\vd400 &37\vd4 &900\vd100 &2200\vd200 \\
{\small\bf B-}$\delta$    & {\small\bf 1.9446} &42\vd5 &800\vd80 &2600\vd400 
&nm &nm &nm &42\vd5 &940\vd90 &3200\vd400 &94\vd8 &1800\vd100 &5900\vd500 
&22\vd2 &960\vd90 &2500\vd200 \\
{\small\bf HeI}            & {\small\bf 2.0587} &6\vd2 &380\vd50 &1100\vd200 
&4\vd1 &750\vd200 &1300\vd400 &7\vd2 &730\vd70 &1300\vd400 &34\vd5 &1400\vd100
&2900\vd400 &9\vd1 &710\vd70 &2200\vd200 \\
{\small\bf B-}$\gamma$    & {\small\bf 2.1655} &27\vd4 &730\vd70 &1500\vd200 
&16\vd1 &1100\vd300 &2000\vd400 &56\vd2 &930\vd90 &2800\vd200 &103\vd6 
&2200\vd100 &3900\vd200 &36\vd1 &1200\vd100 &3000\vd400 \\
{\small\bf AlI}            & {\small\bf 2.1099} &np &np &np &nm &nm &nm &np &np
&np &np &np &np &np &np &np \\
{\small\bf AlI}            & {\small\bf 2.1170} &np &np &np &nm &nm &nm &np &np
&np &np &np &np &np &np &np \\
{\small\bf NaI}      & {\small\bf 2.2062\vline} &np &np &np &-2.8\vd0.5 &nm 
&nm &np &np &np &np &np &np &np &np &np \\
{\small\bf NaI}      & {\small\bf 2.2090\vline} & & & & & & & & & & & & & & &\\
{\small\bf CaI}      & {\small\bf 2.2614~\vline} & & & & & & & & & & & & & & &\\
{\small\bf CaI}      & {\small\bf 2.2631~\vline} &np &np &np &-1.9\vd0.9 &nm 
&nm &np &np &np &np &np &np &np &np &np \\
{\small\bf CaI}      & {\small\bf 2.2657~\vline} & & & & & & & & & & & & & & &\\
{\small\bf $^{12}$CO}      & {\small\bf 2.2935} &np &np &np &-2\vd1 &nm &nm 
&np &np &np &np &np &np &np &np &np \\
{\small\bf $^{12}$CO}& {\small\bf 2.3227\vline} & & & & & & & & & & & & & & &\\
{\small\bf NaI}      & {\small\bf 2.3355\vline} &np &np &np &-5\vd1 &nm &nm 
&np &np &np &np &np &np &np &np &np \\
{\small\bf NaI}      & {\small\bf 2.3386\vline} & & & & & & & & & & & & & & &\\
{\small\bf $^{12}$CO}      & {\small\bf 2.3525} &np &np &np &-7\vd1 &nm &nm 
&np &np &np &np &np &np &np &np &np \\
{\small\bf $^{12}$CO}      & {\small\bf 2.3830} &np &np &np &-2\vd1 &nm &nm 
&np &np &np &np &np &np &np &np &np \\
& & & & & & & & & & & & & & & & \\ 
& & & & & & & & & & & & & & & & \\
& &\multicolumn{3}{c}{\bf SW UMa}&\multicolumn{3}{c}{\bf T Leo}
&\multicolumn{3}{c}{\bf BK Lyn}&\multicolumn{3}{c}{\bf WZ Sge}& & &\\
{\small\bf Line}&$\lambda$&\bf{EW}&\bf{FWHM}&\bf{FWZI}&\bf{EW}&\bf{FWHM}
&\bf{FWZI}&\bf{EW}&\bf{FWHM}&\bf{FWZI}&\bf{EW}&\bf{FWHM}&\bf{FWZI}& & &\\
&{$\mu$\bf{m}}&\bf{\AA}&{\small\bf kms$^{-1}$}&{\small\bf kms$^{-1}$}
&\bf{\AA}&{\small\bf kms$^{-1}$}&{\small\bf kms$^{-1}$}&\bf{\AA}
&{\small\bf kms$^{-1}$}&{\small\bf kms$^{-1}$}&\bf{\AA}&{\small\bf kms$^{-1}$}
&{\small\bf kms$^{-1}$}& & &\\
{\small\bf B-}$\epsilon$  & {\small\bf 1.8174}& nm &1600\vd200& nm & nm & 1200\vd200&nm&nm&600\vd200&nm&nm&nm&nm & & &\\
{\small\bf P-}$\alpha$     & {\small\bf 1.8751} &244\vd9 &1300\vd100
&4500\vd200 &157\vd6 &1600\vd100 & 3700\vd200 &30\vd5 &800\vd80 &1900\vd200
&nm &nm &nm & & &\\
{\small\bf B-}$\delta$    &{\small\bf 1.9446} &128\vd5 &1300\vd100 
&2300\vd200 &93\vd3 &1300\vd100 &4000\vd200 &18\vd3 &850\vd80 &1800\vd200 
&nm &nm &nm & & &\\
{\small\bf HeI}            & {\small\bf 2.0587}&34\vd3 &920\vd90 &2000\vd200
&32\vd2 &980\vd90 &2300\vd200 &5\vd2 &580\vd50 &2200\vd400 &26\vd1 &1600\vd100
&3800\vd200 & & &\\
{\small\bf B-}$\gamma$    & {\small\bf 2.1655} &147\vd4 &1400\vd100 
&3500\vd200 &119\vd2 &1400\vd100 &3500\vd200 &21\vd2 &920\vd150 &2900\vd200
&80\vd1 &2200\vd150 &3400\vd200 & & &\\
{\small\bf AlI}            & {\small\bf 2.1099} &np &np &np &np &np &np &np &np
&np &nm &nm &nm & & &\\
{\small\bf AlI}            & {\small\bf 2.1170} &np &np &np &np &np &np &np &np
&np &nm &nm &nm & & &\\
{\small\bf NaI}      & {\small\bf 2.2062\vline} &np &np &np &np &np &np &np &np
&np &nm &nm &nm & & &\\
{\small\bf NaI}      & {\small\bf 2.2090\vline} & & & & & & & & & & & & & & &\\
{\small\bf CaI}      & {\small\bf 2.2614~\vline} & & & & & & & & & & & & & & &\\
{\small\bf CaI}      & {\small\bf 2.2631~\vline} &np &np &np &np &np &np &np &np
&np &nm &nm &nm & & &\\
{\small\bf CaI}      & {\small\bf 2.2657~\vline} & & & & & & & & & & & & & & &\\
{\small\bf $^{12}$CO}      & {\small\bf 2.2935} &np &np &np &np &np &np &np &np
&np &nm &nm &nm & & &\\
{\small\bf $^{12}$CO}& {\small\bf 2.3227\vline} & & & & & & & & & & & & & & &\\
{\small\bf NaI}      & {\small\bf 2.3355\vline} &np &np &np &np &np &np &np &np
&np &nm &nm &nm & & &\\
{\small\bf NaI}      & {\small\bf 2.3386\vline} & & & & & & & & & & & & & & &\\
{\small\bf $^{12}$CO}      & {\small\bf 2.3525} &np &np &np &np &np &np &np &np
&np &nm &nm &nm & & &\\
{\small\bf $^{12}$CO}      & {\small\bf 2.3830} &np &np &np &np &np &np &np &np
&np &nm &nm &nm  & & &\\
 & & & & & & & & & & & & & & & & \\ 
& & & & & & & & & & & & & & & & \\ 
& & & & & & & & & & & & & & & & \\
& &\multicolumn{3}{c}{\bf DW UMa}&\multicolumn{3}{c}{\bf V1315 Aql}
&\multicolumn{3}{c}{\bf GP Com}& & & &\\
{\small\bf Line}&$\lambda$&\bf{EW}&\bf{FWHM}&\bf{FWZI}&\bf{EW}&\bf{FWHM}
&\bf{FWZI}&\bf{EW}&\bf{FWHM}&\bf{FWZI}& & & & & &\\
&{$\mu$\bf{m}}&\bf{\AA}&{\small\bf kms$^{-1}$}&{\small\bf kms$^{-1}$}
&\bf{\AA}&{\small\bf kms$^{-1}$}&{\small\bf kms$^{-1}$}&\bf{\AA}
&{\small\bf kms$^{-1}$}&{\small\bf kms$^{-1}$}& & & & & &\\
{\small\bf B-}$\epsilon$  & {\small\bf 1.8174}& nm & nm & nm & nm & nm &nm&nm& nm &nm& & & & & &\\
{\small\bf P-}$\alpha$     & {\small\bf 1.8751} &79\vd3 &850\vd30
&1900\vd160 &61\vd8 &1200\vd100 & 3400\vd200 &nm &nm &nm
& & & & & &\\
{\small\bf B-}$\delta$    &{\small\bf 1.9446} &31\vd1 &930\vd150 
&2200\vd200 &32\vd2 &1100\vd100 &2200\vd200 &nm &nm &nm 
& & & & & &\\
{\small\bf HeI}            & {\small\bf 2.0587}&7\vd1 &730\vd50 &1800\vd200
&11\vd1 &820\vd80 &1600\vd200 &200\vd7 &1660\vd70 &4100\vd150 & &
& & & &\\
{\small\bf B-}$\gamma$    & {\small\bf 2.1655} &44\vd1 &1130\vd20 
&3200\vd150&46\vd1 &1400\vd50 &2600\vd200 &51\vd7 &2600\vd400 &4400\vd400
& & & & & &\\
{\small\bf AlI}            & {\small\bf 2.1099} &np &np &np &np &np &np &np &np
&np & & & & & &\\
{\small\bf AlI}            & {\small\bf 2.1170} &np &np &np &np &np &np &np &np
&np & & & & & &\\
{\small\bf NaI}      & {\small\bf 2.2062\vline} &np &np &np &np &np &np &np &np
&np & & & & & &\\
{\small\bf NaI}      & {\small\bf 2.2090\vline} & & & & & & & & & & & & & & &\\
{\small\bf CaI}      & {\small\bf 2.2614~\vline} & & & & & & & & & & & & & & &\\
{\small\bf CaI}      & {\small\bf 2.2631~\vline} &np &np &np &np &np &np &np &np
&np & & & & & &\\
{\small\bf CaI}      & {\small\bf 2.2657~\vline} & & & & & & & & & & & & & & &\\
{\small\bf $^{12}$CO}      & {\small\bf 2.2935} &np &np &np &np &np &np &np &np
&np & & & & & &\\
{\small\bf $^{12}$CO}& {\small\bf 2.3227\vline} & & & & & & & & & & & & & & &\\
{\small\bf NaI}      & {\small\bf 2.3355\vline} &np &np &np &np &np &np &np &np
&np & & & & & &\\
{\small\bf NaI}      & {\small\bf 2.3386\vline} & & & & & & & & & & & & & & &\\
{\small\bf $^{12}$CO}      & {\small\bf 2.3525} &np &np &np &np &np &np &np &np
&np & & & & & &\\
{\small\bf $^{12}$CO}      & {\small\bf 2.3830} &np &np &np &np &np &np &np &np
&np & & & & & &\\
 & & & & & & & & & & & & & & & & \\ 
\end{tabular*}
}
}
\end{center}

\renewcommand{\arraystretch}{1.0}
\label{tab:lines}
\end{table*}

\clearpage
\newpage

\newcounter{junk}
\setcounter{junk}{5}
\begin{table*}
\caption[]{Wavelengths and equivalent widths (in \AA ngstroms) of the most
prominent lines visible in the IR spectra of the surveyed dwarf stars. 
The line identifications have been based upon the list presented
by \protect\scite{dhillon95a} and references therein. 
The wavelengths given for the $^{12}$CO lines refer to the band-heads. 
Equivalent widths 
for the H$_2$0 band were measured between 2.29 and 2.44 $\mu$m. 
The vertical bars following lines of similar wavelength indicate that the 
equivalent width measurements apply to the entire blend. The two-letter codes
indicate that a line was either 
not present (np) or present but not measurable (nm).
Note that the 
values indicated by asterisks are subject to unkown systematic errors
due to difficulties in continuum normalisation.} 

\newcommand{\vd}{$\pm$}
\renewcommand{\arraystretch}{1.3}
\boldmath{
\hspace{-1.5cm}
{\scriptsize
\begin{tabular*}{20.0cm}{@{\extracolsep{-2.15mm}}llccccccccccccc}
& & & & & & & & & & & & & & \\
{\small\bf Line}&{$\lambda$ ($\mu$\bf{m})}
&\multicolumn{1}{c}{\bf Gl764.1A}&\multicolumn{1}{c}{\bf Gl775}
&\multicolumn{1}{c}{\bf Gl764.1B}&\multicolumn{1}{c}{\bf Gl154}
&\multicolumn{1}{c}{\bf Gl763}&\multicolumn{1}{c}{\bf Gl229}
&\multicolumn{1}{c}{\bf Gl806}
&\multicolumn{1}{c}{\bf Gl436}&\multicolumn{1}{c}{\bf Gl748AB}
&\multicolumn{1}{c}{\bf Gl402}&\multicolumn{1}{c}{\bf Gl866AB}
&\multicolumn{1}{c}{\bf Gl473AB}&\multicolumn{1}{c}{\bf Gl65AB}\\
& &K2\Roman{junk} &K5\Roman{junk} &K7\Roman{junk} &M0\Roman{junk} 
&M0\Roman{junk} &M1\Roman{junk} &M2\Roman{junk} &M3\Roman{junk} 
&M3.5\Roman{junk} &M4\Roman{junk} &M5\Roman{junk} &M5.5\Roman{junk} 
&M6\Roman{junk} \\ 
 & & & & & & & & & & & & & & \\ 
{\small\bf AlI}      & {\small\bf 2.1099\vline} &nm &-4.3\vd0.5 &-3.7\vd0.6 
&-3.1\vd0.6 &-2.1\vd0.4 &-1.3\vd0.3 &-1.3\vd0.4 &-0.8\vd0.4 &nm &nm 
&nm &nm &nm \\
{\small\bf AlI}      & {\small\bf 2.1170\vline} & & & & & & & & & & & & & \\
{\small\bf NaI}      & {\small\bf 2.2062~\vline} &nm &-2.1\vd0.4 &-2.9\vd0.4 
&-3.2\vd0.6 &-3.0\vd0.3 &-4.2\vd0.4 &-3.8\vd0.4 &-4.9\vd0.4 &-3.1\vd0.4 
&-6.7\vd0.5 &-6.6\vd0.6 &-5.8\vd0.3 &-6.5\vd0.4 \\
{\small\bf NaI}      & {\small\bf 2.2090~\vline} & & & & & & & & & & & & & \\
{\small\bf CaI}      & {\small\bf 2.2614\vline} & & & & & & & & & & & & & \\
{\small\bf CaI}      & {\small\bf 2.2631\vline} &-1.2\vd0.4 &-2.9\vd0.4 
&-3.1\vd0.5 &-3.5\vd0.5 &-2.9\vd0.5 &-6.4\vd0.4 &-4.4\vd0.4 &-4.3\vd0.4 
&-2.9\vd0.4 &-4.6\vd0.5 &-6.0\vd0.7 &-3.5\vd0.3 &-2.3\vd0.3 \\
{\small\bf CaI}      & {\small\bf 2.2657\vline} & & & & & & & & & & & & & \\
{\small\bf H$_2$0}         & {\small\bf 2.2900} &-26\vd2 &-35\vd2 &-80\vd2 
&-61\vd3 &-62\vd2 &-72\vd2$^*$ &-152\vd2 &-194\vd1 &-222\vd2 &-249\vd2 
&-346\vd2 &-315\vd1 &-235\vd2 \\
{\small\bf $^{12}$CO}      & {\small\bf 2.2935} &-3.3\vd0.6 &-1.9\vd0.6 
&-3.6\vd0.6 &-9\vd1 &-3.1\vd0.7 &-16\vd1$^*$ &-3.9\vd0.6 &-2.0\vd0.6 
&-3.8\vd0.5 &-4\vd1 &-6.1\vd0.8 &-4.0\vd0.4 &-3.9\vd0.5 \\
{\small\bf $^{12}$CO}& {\small\bf 2.3227\vline} & & & & & & & & & & & & & \\ 
{\small\bf NaI}      & {\small\bf 2.3355\vline} &-4\vd1 &-5.1\vd0.7 
&-6.5\vd0.8 &-14\vd1 &-4.8\vd0.9 &-20\vd1$^*$ &-7.2\vd0.8 &-6.9\vd0.7 
&-9.0\vd0.7 &-11\vd1 &-12\vd1 &-9.8\vd0.6 &-8.7\vd0.6 \\
{\small\bf NaI}      & {\small\bf 2.3386\vline} & & & & & & & & & & & & & \\
{\small\bf $^{12}$CO}      & {\small\bf 2.3525} &-2.0\vd0.8 &-4.4\vd0.8 
&-4.1\vd0.9 &-14\vd1 &-1.6\vd0.9 &-18\vd1$^*$ &-3.1\vd0.8 &-4.7\vd0.7 
&-5.2\vd0.7 &-5\vd1 &-7.5\vd1 &-6.1\vd0.6 &-5.5\vd0.8 \\
{\small\bf $^{12}$CO}      & {\small\bf 2.3830} &-0.5\vd0.9 &-2.4\vd0.8 
&-2\vd1 &-11\vd1 &-1.5\vd0.8 &-11\vd1$^*$ &-1.6\vd0.7 &-3.5\vd0.7 
&-2.0\vd0.8 &-2\vd1 &-1.2\vd0.9 &-2.1\vd0.5 &-1.2\vd0.6 \\
 & & & & & & & & & & & & & & \\ 
\end{tabular*}
}
}
\renewcommand{\arraystretch}{1.0}
\label{tab:seclines}
\end{table*}

\begin{table*}
\newcommand{\vd}{$\pm$}
\caption[]{Upper limits to the contributions of the secondary star to the
 total K-band flux.}
\boldmath{
{\bf
\begin{tabular}{lccc}
 & & &\\
\multicolumn{1}{l}{Object} &
\multicolumn{1}{c}{Secondary star} &
\multicolumn{1}{c}{Error} &
\multicolumn{1}{c}{Spectral-type} \\
& \multicolumn{1}{c}{contribution} &
& \multicolumn{1}{c}{template used} \\
 & & & \\
VY~Scl    &  90\% & 10\% & M4V   \\
DW~UMa    &  35\% &  5\% & M4V   \\
V1315~Aql &  30\% &  5\% & M4V   \\
UU~Aqr    &  55\% &  5\% & M4V   \\
BK~Lyn    &  50\% &  5\% & M5V   \\
LY~Hya    & 100\% & 10\% & M5V   \\
YZ~Cnc    &  20\% &  5\% & M5V   \\
SW~UMa    &  20\% &  5\% & M5.5V \\
T~Leo     &  30\% &  5\% & M5.5V \\
WZ~Sge    &  10\% &  5\% & M5.5V \\
 & & & \\
 & & & \\
\end{tabular}
}
}
\label{tab:contrib}
\end{table*}

\end{document}